\documentclass[aps,
prd,
reprint,
twocolumn,
tightenlines,
amsmath,
nofootinbib,
amssymb]{revtex4-2}
\usepackage{bm,braket,color,comment,graphicx,here,multirow,orcidlink,slashed,tikz}
\usepackage{hyperref}
\usetikzlibrary{arrows.meta, decorations.pathmorphing}
\usepackage[compat=1.1.0]{tikz-feynhand}
\usetikzlibrary{arrows}
\def\beq{\begin{equation}}
\def\eeq{\end{equation}}
\def\bea{\begin{eqnarray}}
\def\eea{\end{eqnarray}}
\def\nn{\nonumber}
\hypersetup{hidelinks}
\begin{document}
\title{Polarized observables in \boldmath{$\tau \to VV \nu_{\tau}$} decays}
\author{\orcidlinki{Vindhyawasini Prasad}
{0000-0001-7395-2318}}
\email{Contact author: vindy@jlu.edu.cn}
\affiliation{
Center for Theoretical Physics and College of Physics, Jilin University, Changchun, 130012, People's Republic of China, China}
\author{\orcidlinki{Hiroyuki Umeeda}{0000-0003-0461-0976}}
\email{Contact author: umeeda@jlu.edu.cn}
\affiliation{
Center for Theoretical Physics and College of Physics, Jilin University, Changchun, 130012, People's Republic of China, China}
\date{\today}
\begin{abstract}
We study polarization observables in $\tau^- \to V_1^- V_2^0 \nu_\tau$ decays for $(V_1^-, V_2^0)=(\rho^-, \rho^0), (\rho^-, \omega), (K^{*-}, \rho^0), (K^{*-}, \omega), (\rho^-, \bar{K}^{*0})$ using the helicity formalism. While the obtained branching fraction for $\tau^-\to \rho^-\rho^0\nu_\tau$ is consistent with that of angular momentum algebra method, the results for the other four channels are comparable with (smaller than) those from the effective chiral model (the aforementioned method). Our results for the longitudinal polarization fractions $f_L$ indicate that nonnegligible corrections to the endpoint limit ($f_L = 1/3$), a consequence of Lorentz symmetry, are present at the scale of tau mass. Future measurements of $f_L$ at Belle II for $\tau^- \to V_1^- V_2^0 \nu_\tau$ decays will provide crucial information on polarization patterns observed in nonleptonic charm decays.
\end{abstract}
\maketitle
\noindent\textit{Introduction.} Processes involving two vector mesons in the final state provide a notable testing ground for polarization studies. Numerous works have been performed in $B \to VV$ decays, while measurements in $D \to VV$ decays \cite{MARK-III:1991fvi,FOCUS:2007ern,BESIII:2021raf,BESIII:2026dwz} have also been carried out. The $VV$ final state is characterized by two transverse and one longitudinal polarization mode, whose fractions are determined by strong-interaction dynamics. In $B \to VV$ decays, longitudinal polarization typically dominates due to quark-level helicity conservation \cite{Ali:1978kn}. However, measurements \cite{BaBar:2004uwv,Belle:2005lvd,BaBar:2006ttd,BaBar:2007bpi,BaBar:2008lan,LHCb:2013nlu,Belle:2013vat,LHCb:2014xzf} indicate that $B_{(s)} \to K^*\phi$ decays deviate from this expectation, referred to as the {\lq polarization puzzle\rq}. Possible interpretations for this puzzle include rescattering, penguin annihilation, or new physics (see Ref.~\cite{ParticleDataGroup:2024cfk} for a review and references therein).
\par
In $D \to V_1V_2$ decays, the longitudinal polarization fraction $(f_L)$ is estimated to be $0.31\leq f_L \leq 0.49$ by the factorization approach \cite{Cheng:2024hdo}. Another method that relies on kinematical restrictions from Lorentz symmetry \cite{Hiller:2013cza,Hiller:2021zth,Zwicky:2013eda} predicts $f_L\simeq 1/3$, provided that an approximation of $m_{D}\simeq m_{V_1}+m_{V_2}$ ($m$ with an index represents mass of each particle) works properly. However, those results (see also Refs.~\cite{Cao:2023csx, Ou-Yang:2026fyc}) cannot simultaneously reproduce all measured $f_L$ in experiments \cite{MARK-III:1991fvi,FOCUS:2007ern,BESIII:2021raf,BESIII:2026dwz}; puzzling patterns are also present in $D\to VV$ decays, as well as $B$-meson case. In this context, an independent polarization study in $\tau$ decays is highly desirable, especially to clarify whether the predictions from Lorentz symmetry can be regarded as the lowest-order approximation.
\par
In this work, differential decay widths in $\tau\to VV\nu_\tau$ are discussed for evaluating longitudinal polarization fractions. Observables of interest can be analyzed by means of the helicity formalism \cite{Jacob:1959at} (see Ref.~\cite{Richman:1984gh} for a review), which enables us to extract angular dependence in a way analogous to $B\to VV$ \cite{Dunietz:1990cj, Kramer:1991xw, Dighe:1995pd}, $\bar{B}\to D^*\tau\bar{\nu}_\tau$ \cite{Korner:1989qb,Fajfer:2012vx}, and $\Xi^0\to \Sigma^+\ell\bar{\nu}_\ell$ \cite{Kadeer:2005aq} decays. For $\tau\to VV\nu_\tau$ decays, there exist seven nonvanishing polarized configurations for vector mesons in the standard model (SM), to be contrasted with $D\to VV$ decays, where only three modes are present (see Appendix A, especially for discussions around Eqs.~(\ref{Eq:Helconf1})-(\ref{Eq:VVvanish})). In order to estimate contributions of lowest-lying vector, axial vector, and pseudoscalar mesons, we adopt generalized hidden local symmetry (GHLS) \cite{Bando:1987ym} Lagrangian, an extension of the hidden local symmetry (HLS) \cite{Bando:1984ej,Bando:1987br,Harada:2003jx} (see Refs.~\cite{Ecker:1988te,Ecker:1989yg,Li:1995aw,Li:1995tv,Li:1996md,Li:1996ks,Kimura:2012bwp,Kimura:2016xnx,Dai:2018thd,Dai:2018zki,Dai:2020vfc} for other methods). Contributions of excited intermediate mesons ($M^\prime$) are included via $V_1V_2M^\prime$ and $WM^\prime$ couplings with Lorentz structure identical to the ones in GHLS. Numerical results are given for branching ratios and $f_L$ with Cabibbo-favored (CF) and Cabibbo-suppressed (CS) processes.
\\
\\
\noindent
\textit{Angular distributions.} $\tau$ decays into two mesons in the final state were discussed, {\it e.g.}, in Refs.~\cite{Kuhn:1992nz,Decker:1992jy,Decker:1992kj,Kuhn:1996dv}, which our analysis largely follows in the helicity formalism. The hadronic helicity amplitudes (HAs) for $\tau^-\to V_1^-V_2^0\nu_\tau$ decays ($V_i$ for $i=1, 2$ represents a vector meson) 
are defined via a matrix element of the quark $V-A$ current ($J^\mu_q$ for $q=d, s$), projected onto  polarization states as given below:
\bea
&H_{\lambda_1 \lambda_2}^{(m)}=
H^\mu_{\lambda_1\lambda_2}\epsilon_\mu(m),\quad
J^\mu_q=\bar{q}\gamma^\mu(1-\gamma^5)u,
&
\label{Eq:heldefeq}\\
&H^\mu_{\lambda_1\lambda_2}=
\braket{V_1^-(p_1, \lambda_{1})
V_2^0(p_2, \lambda_{2})|J^\mu_q|0},&
\eea
where $p_i$ and $\lambda_i$ represent the momentum and helicity of $V_i$, respectively, and similar notations are used in what follows. A proper basis of the polarization vector, $\epsilon_\mu(m)$ with intermediate helicity indices of $m=s, \pm, 0$ in Eq.~(\ref{Eq:heldefeq}), is given by Eqs.~(\ref{Eq:A1})-(\ref{Eq:A3}). Furthermore, the size of a momentum of either $V_1^-$ or $V_2^0$ in the rest frame of the hadronic system is
\bea
|\bm{p}|=\sqrt{\lambda(Q^2, m_{V_1}^2, m_{V_2}^2)}/\left(2\sqrt{Q^2}\right),
\label{Eq:labelofp}
\eea
with $Q=p_1+p_2$ and the K\"{a}ll\'{e}n function defined as $\lambda(a, b, c)=a^2+b^2+c^2-2ab-2bc-2ca$.
\par
Helicity angles in $\tau^-\to V_1^-V_2^0\nu_\tau$ decays are illustrated in Fig.~\ref{Fig:1}, taking $(V_1^-, V_2)=(\rho^-, \omega)$ as an example: $\theta$ is associated with $\nu_\tau$ direction in the rest frame of hadronic system, while $\theta_{i}~(\phi_i)$ is a polar (azimuthal) angle defined in the rest frame of $V_i$ for $i=1, 2$. Although generic distributions expressed by multiple variables can be considered, we restrict ourselves to the polar angle distributions in the present work. Integrating over variables other than $\cos\theta_{i}$, the differential decay rate for unpolarized $\tau^-$ is \cite{SupplementalMaterial}
\bea
\frac{1}{\Gamma}
\displaystyle\frac{d\Gamma}
{d\cos\theta_i}
&=&
\frac{3}{4}\sin^2\theta_i\left[1-f_L^{(V_i)}\right]
+\frac{3}{2}\cos^2\theta_if_L^{(V_i)},
\label{Eq:ResAngfinalfin2}\\
\Gamma&=&\frac{G_F^2|V_{uq}|^2m_\tau}{192\pi^3}K.\label{Eq:ResAngfinalfin3}
\eea
In the above relation, $G_F$ denotes the Fermi constant, 
$V_{uq}~(q=d, s)$ represents the Cabibbo-Kobayashi-Maskawa (CKM) matrix \cite{Cabibbo:1963yz,Kobayashi:1973fv} element, and $m_\tau$ is $\tau$ lepton mass. In Eq.~(\ref{Eq:ResAngfinalfin2}), the longitudinal polarization fractions for two vector mesons are \cite{SupplementalMaterial},
\bea
f_L^{(V_1)}=
\displaystyle\sum_{\lambda_2}
K_{0\lambda_2}/K,\qquad
f_L^{(V_2)}=
\displaystyle\sum_{\lambda_1}
K_{\lambda_10}/K,\label{Eq:FLdefrelations}
\eea
where $K$ and $K_{\lambda_1\lambda_2}$ in Eqs.~(\ref{Eq:ResAngfinalfin3}) and (\ref{Eq:FLdefrelations}) are the objects that quadratically depend on hadronic HAs, defined by Eqs.~(\ref{Eq:Kequationfirst}) and (\ref{Eq:Keqationsec}).
\begin{figure}[t]
\begin{tikzpicture}[scale=0.50]
\coordinate[] (A) at (0,0);
\coordinate[] (B) at (4.5,0);
\coordinate[] (C) at (-4.5,0);
\node[] (hoge) at (0, -0.30) {\tiny $W^{- *}$};
\node[] (hoge) at (+4.5, 0.43) {\tiny$V_1^-$};
\node[] (hoge) at (-4.5, 0.43) {\tiny$V_2^0$};
\node[] (hoge) at (6.58, 1.02) {\tiny$P^-$};
\node[] (hoge) at (2.82, -1.05) {\tiny$P^0$};
\node[] (hoge) at (-6.50, -1.07) {\tiny$P^+$};
\node[] (hoge) at (-3.7, 1.0) {\tiny$\bm{p}_{P^-}+\bm{p}_{P^0}$};
\node[] (hoge) at (1.4, 0.72) {\tiny$-\bm{p}_{\nu_\tau}$};
\node[] (hoge) at ( 3.52, -0.01) {\tiny$\cdots$};
\node[] (hoge) at (-3.48, -0.01) {\tiny$\cdots$};
\node[] (hoge) at (0.7, 0.19) {\tiny$\theta$};
\node[] (hoge) at (5.6, 0.21) {\tiny$\theta_1$};
\node[] (hoge) at (1.82, -1.65) {\tiny$\phi_1$};
\node[] (hoge) at (-5.56, -0.24) {\tiny$\theta_2$};
\node[] (hoge) at (-1.70, 1.69) {\tiny$\phi_2$};
\draw (0.5, 0) to [out=30,in=30] (0.4, 0.22);
\draw (5.2, 0) to [out=30,in=30] (5.1, 0.345);
\draw (-5.2, 0) to [out=150,in=150] (-5.1, -0.345);
\draw (2.1, -1.5) to [out=-80,in=-80] (1.74, -1.2);
\draw (-2.1, 1.5) to [out=80,in=80] (-1.74, 1.2);
\draw[gray, thick, line width=2pt]  (-1.7, 1.3) -- ( 2.73, 1.3);
\draw[gray, dotted, line width=2pt] (2.73, 1.3) -- ( 3.2, 1.3);
\draw[gray, thick, line width=2pt]  (+1.7,-1.3) -- (-2.73,-1.3);
\draw[gray, dotted, line width=2pt]  (-2.73,-1.3) -- (-3.2,-1.3);
\draw[gray, dotted, line width=2pt] (-3.2,-1.3) -- (-3.2+0.75, 0);
\draw[gray, thick, line width=2pt]  (-3.2+0.75, 0) -- (-3.2+1.5, 1.3);
\draw[gray, thick, line width=2pt]  (+1.7,-1.3) -- ( 1.7+0.75, 0);
\draw[gray, dotted, line width=2pt]  (1.7+0.75,0) -- ( 3.2, 1.3);
\draw[blue, thick, line width=2pt]  (2.8+4.6, 1.6) -- ( 2.8, 1.6);
\draw[blue, thick, line width=2pt]  (2.1+4.6,-1.6) -- ( 2.1,-1.6);
\draw[blue, thick, line width=2pt]  (+2.1+4.6,-1.6) -- ( 2.8+4.6, 1.6);
\draw[blue, thick, line width=2pt]  (+2.1,-1.6) -- ( 2.8, 1.6);
\draw[red, thick, line width=2pt]  (2.8+4.6 -9.5, 1.6) -- ( 2.8    -9.5, 1.6);
\draw[red, thick, line width=2pt]  (2.1+4.6 -9.5,-1.6) -- ( 2.1    -9.5,-1.6);
\draw[red, thick, line width=2pt]  (+2.1+4.6-9.5,-1.6) -- ( 2.8+4.6-9.5, 1.6);
\draw[red, thick, line width=2pt]  (+2.1    -9.5,-1.6) -- ( 2.8    -9.5, 1.6);
\draw[gray, thick]  ( 2.35,0) -- ( 3.2, 0);
\draw[gray, thick]  ( 3.8, 0) -- ( 4.5, 0);
\draw[gray, thick]  (-2.35,0) -- (-3.2, 0);
\draw[gray, thick]  (-3.8, 0) -- (-4.5, 0);
\draw[->,>=stealth] (0,0)--(1.1, 0.61);
\draw[->,>=stealth] (0,0)--( 2.5,0);
\draw[->,>=stealth] (0,0)--(-2.5,0);
\draw[->,>=stealth] (4.5, 0)--(7.5,0);
\draw[->,>=stealth] (4.5, 0)--(4.5-2*0.75, -1.1*0.75);
\draw[->,>=stealth] (4.5, 0)--(4.5+2*0.85, +1.1*0.85);
\draw[->,>=stealth] (-4.5, 0)--(-7.5,0);
\draw[->,>=stealth] (-4.5, 0)--(-4.5+2*0.75, 1.1*0.75);
\draw[->,>=stealth] (-4.5, 0)--(-4.5-2*0.85, -1.1*0.85);
\fill (A) circle[radius=0.1];
\fill (B) circle[radius=0.1];
\fill (C) circle[radius=0.1];
\end{tikzpicture}
\vspace{-4mm}
\caption{Definitions of angles for $W^{-*}\to V_1^-(\to P^- P^0)V_2^0(\to P^+P^-P^0)$, where $P$ represents a pseudoscalar, in unpolarized $\tau^-\to V_1^- V_2^0\nu_\tau$ decays. Each plane is defined in the rest frame of the parent particle with ellipses indicating that different Lorentz frames are considered.}
\label{Fig:1}
\vspace{-2mm}
\end{figure}
\par
There is a couple of differences between $f_L$ in $\tau^-\to V_1^-V_2^0\nu_\tau$ and $D\to V_1V_2$ (or similarly $B\to V_1V_2$) given as follows: ($\alpha$) For $\tau^-\to V_1^-V_2^0\nu_\tau$ decays, each $K$ in Eq.~(\ref{Eq:Kequationfirst}) are defined by an integral over $Q^2$ with the specific weight factor in Eq.~(\ref{Eq:Keqationsec}), and an additional coefficient of $m_\tau^2/Q^2$ in Eq.~(\ref{Eq:expresfuncI}) for $I^{(-1/2)}_{\lambda_1\lambda_2}$, which have kinematical dependences from three-body phase space combined with the leptonic HAs. Such an involved integral is absent for two-body $D\to V_1V_2$ decays.
($\beta$)
The two objects of $f_L^{(V_1)}$ and $f_L^{(V_2)}$ in $\tau^-\to V_1^-V_2^0\nu_\tau$ decays are to be distinguished in general, unlike $D\to V_1V_2$ decays. 
\\
\\
\noindent
\textit{Hadronic helicity amplitudes.} Here, the longitudinal polarization fractions and the associated angular distributions are analyzed for $\tau^-\to V^-_1(p_1, \epsilon_1)V_2^0(p_2, \epsilon_2)\nu_\tau$ decays, where $\epsilon_i$ represents a polarization vector of $V_i$. Relevant diagrams are displayed in Fig.~\ref{Fig:2}. For (a), (b) and (c) in the figure, intermediate resonances for both lowest-lying state and excited state(s) are considered. A certain basis for vector-meson polarizations in the rest frame of the hadronic system is given by Eqs.~(\ref{Eq:Vpol1})-(\ref{Eq:Vpol2}). 
\par
For $(V_1^-, V_2^0)=(\rho^-, \rho^0)$, nonvanishing contributions arise only from the vector current, and thus 
contributions of Fig.~\ref{Fig:2}(b), (c), (d) are absent due to $G$-parity conservation \footnote{Since $\rho^-$ and $\omega$ are respectively even and odd in $G$-parity, they multiplicatively give a $G$-parity odd final state.}. Similarly, Fig.~2(a) vanishes for the case of $(V_1^-, V_2^0)=(\rho^-, \omega)$. As to CS channels, all types of contributions in Fig.~\ref{Fig:2} are nonzero in general. The hadronic HA reads,
\bea
&&
H^{(m)}_{\lambda_1\lambda_2}=
\mathcal{H}^{(V)}_{V_1V_2}+
\mathcal{H}^{(A)}_{V_1V_2}+
\mathcal{H}^{(P)}_{V_1V_2}+
\mathcal{H}^{(\textrm{NR})}_{V_1V_2}.
\label{Eq:heldefi}
\eea
In the above relation, $(V), (A), (P)$ denote the 
vector, axial vector and pseudoscalar intermediate resonances, while $(\textrm{NR})$ corresponds to the nonresonant contribution. Each term in Eq.~(\ref{Eq:heldefi}) is given by
\begin{figure}[t]
\begin{tikzpicture}[scale=0.42]
\begin{feynhand}
    \vertex [particle] (i1) at (-3,1) {$\tau^-$};
    \vertex [particle] (f1) at (+3,1) {$\nu_{\tau}$};
    \vertex (w1) at (0, 0);
    \propag [fermion] (i1) to (w1);
    \propag [fermion] (w1) to (f1);  
    \fill (w1) circle[radius=0.1];
    \vertex (w2) at (0.0, -1.2);
    \propag [boson] (w1) to node[left, pos=0.5] {$W^-$} (w2);
    \fill   (w2) circle[radius=0.1];
    \vertex (w3) at (0.0, -2.4);
    \propag [boson] (w2) to node[left, pos=0.5] {$\rho^-/K^{*-}$}(w3);
    \fill   (w3) circle[radius=0.1];
    \vertex [particle] (f2) at (3,-1.5) {$V_1^-$};
    \vertex [particle] (f3) at (3,  -3) {$V_2^0$};
    \propag [boson] (w3) to (f2);
    \propag [boson] (w3) to (f3);
    \vertex [particle] (i0) at (0, -3.75) {(a)};
\end{feynhand}
\end{tikzpicture}
\hspace{11mm}
\begin{tikzpicture}[scale=0.42]
\begin{feynhand}
    \vertex [particle] (i1) at (-3,1) {$\tau^-$};
    \vertex [particle] (f1) at (+3,1) {$\nu_{\tau}$};
    \vertex (w1) at (0, 0);
    \propag [fermion] (i1) to (w1);
    \propag [fermion] (w1) to (f1);  
    \fill (w1) circle[radius=0.1];
    \vertex (w2) at (0.0, -1.2);
    \propag [boson] (w1) to node[left, pos=0.5] {$W^-$} (w2);
    \fill   (w2) circle[radius=0.1];
    \vertex (w3) at (0.0, -2.4);
    \propag [boson] (w2) to node[left, pos=0.5] {$a_1^-/K_{1A}^{-}$}(w3);
    \fill   (w3) circle[radius=0.1];
    \vertex [particle] (f2) at (3.0,-1.5) {$V_1^-$};
    \vertex [particle] (f3) at (3,  -3) {$V_2$};
    \propag [boson] (w3) to (f2);
    \propag [boson] (w3) to (f3);
    \vertex [particle] (i0) at (0, -3.75) {(b)};
\end{feynhand}
\end{tikzpicture}
\\
\begin{tikzpicture}[scale=0.42]
\begin{feynhand}
    \vertex [particle] (i1) at (-3,1) {$\tau^-$};
    \vertex [particle] (f1) at (+3,1) {$\nu_{\tau}$};
    \vertex (w1) at (0, 0);
    \propag [fermion] (i1) to (w1);
    \propag [fermion] (w1) to (f1);  
    \fill (w1) circle[radius=0.1];
    \vertex (w2) at (0.0, -1.2);
    \propag [boson] (w1) to node[left, pos=0.5] {$W^-$} (w2);
    \fill   (w2) circle[radius=0.1];
    \vertex (w3) at (0.0, -2.4);
    \propag [scalar] (w2) to node[left, pos=0.5] {$\pi^-/K^-$}(w3);
    \fill   (w3) circle[radius=0.1];
    \vertex [particle] (f2) at (3,-1.5) {$V_1^-$};
    \vertex [particle] (f3) at (3,  -3) {$V_2^0$};
    \propag [boson] (w3) to (f2);
    \propag [boson] (w3) to (f3);
    \vertex [particle] (i0) at (0, -3.75) {(c)};
\end{feynhand}
\end{tikzpicture}
\hspace{10.5mm}
\begin{tikzpicture}[scale=0.42]
\begin{feynhand}
    \vertex [particle] (i1) at (-3,1) {$\tau^-$};
    \vertex [particle] (f1) at (+3,1) {$\nu_{\tau}$};
    \vertex (w1) at (0, 0);
    \propag [fermion] (i1) to (w1);
    \propag [fermion] (w1) to (f1);  
    \fill (w1) circle[radius=0.1];
    \propag [boson] (w1) to node[left, pos=0.5] {$W^-$} (w3);
    \vertex (w3) at (0.0, -2.4);
    \fill   (w3) circle[radius=0.1];
    \vertex [particle] (f2) at (3,-1.5) {$V_1^-$};
    \vertex [particle] (f3) at (3,  -3) {$V_2^0$};
    \propag [boson] (w3) to (f2);
    \propag [boson] (w3) to (f3);
    \vertex [particle] (i0) at (0, -3.75) {(d)};
\end{feynhand}
\end{tikzpicture}
\hspace{11.2mm}
\vspace{-3mm}
\caption{Diagrams for $\tau^-\to V^-_1V_2^0 \nu_\tau$ decays for
(a) a vector meson, 
(b) an axial-vector meson, 
(c) a pseudoscalar meson, with
(d) a nonresonant coupling,
for both CF/CS channels.
}
\label{Fig:2}
\end{figure}
\bea
\mathcal{H}^{(V)}_{V_1V_2}&=&
\displaystyle\sum_{\mathcal{V}}
C_{V_1V_2\mathcal{V}}^{(V)}T^{(V)}_{\mathcal{V}},
\label{Eq:HVexpre1}\\
\mathcal{H}^{(A)}_{V_1V_2}&=&
\displaystyle\sum_{\mathcal{A}}
C_{V_1V_2\mathcal{A}}^{(A)}T^{(A)}_{\mathcal{A}},
\label{Eq:HVexpre2}\\
\mathcal{H}^{(P)}_{V_1V_2}&=&
\displaystyle\sum_{\mathcal{P}}
C_{V_1V_2\mathcal{P}}^{(P)}T^{(P)},
\label{Eq:HVexpre3}\\
\mathcal{H}^{(\textrm{NR})}_{V_1V_2}&=&
G_{V_1V_2W} T^{(\textrm{NR})}
\label{Eq:NRcoeff},
\eea
where the sums in Eqs.~(\ref{Eq:HVexpre1})-(\ref{Eq:HVexpre3}) run over intermediate states for both lowest-lying and excited resonances: in the case of the CF channels, the summation is taken for
\bea
\mathcal{V}&=&\rho(770),~\rho(1450),~\rho(1700),\label{Eq:Exci1}\\
\mathcal{A}&=&a_1(1260),~a_1(1640),\\
\mathcal{P}&=&\pi,~\pi(1300),~\pi(1800)\label{Eq:Exci3},
\eea
while the contributions relevant to the CS modes are
\bea
\mathcal{V}&=&K^*(892),~K^*(1410),~K^*(1680),\label{Eq:Exci4}\\
\mathcal{A}&=&K_{1A},\\
\mathcal{P}&=&K,~K(1460).\label{Eq:Exci6}
\eea
Detailed formulae for $C_{V_1V_2 M}^{(i)}~(i=V, A, P)$, $G_{V_1V_2W}$, $T^{(V)}_{\cal V}$,
$T^{(A)}_{\cal A}$,
$T^{(P)}$, and $T^{(\textrm{NR})}$ in Eqs.~(\ref{Eq:HVexpre1})-(\ref{Eq:NRcoeff}) 
are given in Appendix D.
\par
For Fig.~\ref{Fig:2}(d) and the lowest-lying contributions in (a), (b), and (c),
(G)HLS \cite{Kaiser:1990yf} is adopted for our main results. Alternatively, Fig.~\ref{Fig:2}(b) can be also  extracted from Ref.~\cite{Kuhn:2006nw} for TAUOLA: In that work, $\tau\to 5\pi\nu_\tau$ is evaluated for two types of diagrams, one of which includes the axial-vector meson as an intermediate state. Since $\tau\to \rho\omega\nu_\tau$ is a subdiagram of that process, the hadronic HA for this channel is inferred from Ref.~\cite{Kuhn:2006nw} up to sign relative to the GHLS contributions. Combining the $W^-a_1^-$ and $a_1\rho\omega$ couplings, the overall constant of this contribution is parametrized as $c_0$ in Eq.~(\ref{Eq:Tauoexp1}).
\par
A certain care must be taken for different expressions of $T^{(A)}_{\rm GHLS}$ in Eq.~(\ref{Eq:AxialGHLS}) and $T^{(A)}_{\rm TAUOLA}$ in Eq.~(\ref{Eq:Tauoexp2}), which originate from distinct $a_1\rho\omega$ vertex structures. For the GHLS case, interaction in the form of $T^{(A)}_{\rm TAUOLA}$ does not appear for parity and charge conjugation invariant Lagrangian with hermiticity \cite{Kaiser:1990yf} (see Ref.~\cite{Harvey:2007ca}, giving the same Lorentz structure of $a_1\rho\omega$ vertex as \cite{Kaiser:1990yf}).  In particular, difference in proportionality to momenta read \footnote{In Ref.~\cite{Kuhn:2006nw}, $a_1\rho\omega$ interaction structures that include the $p_1-p_2$ factor are also introduced. However, the one proportional to $(p_1-p_2)^0$ is adopted in that reference for simplicity.}, 
\begin{figure}[tbp]
  \centering
  \begin{minipage}{\linewidth}
    \centering
    \includegraphics[width=\linewidth]{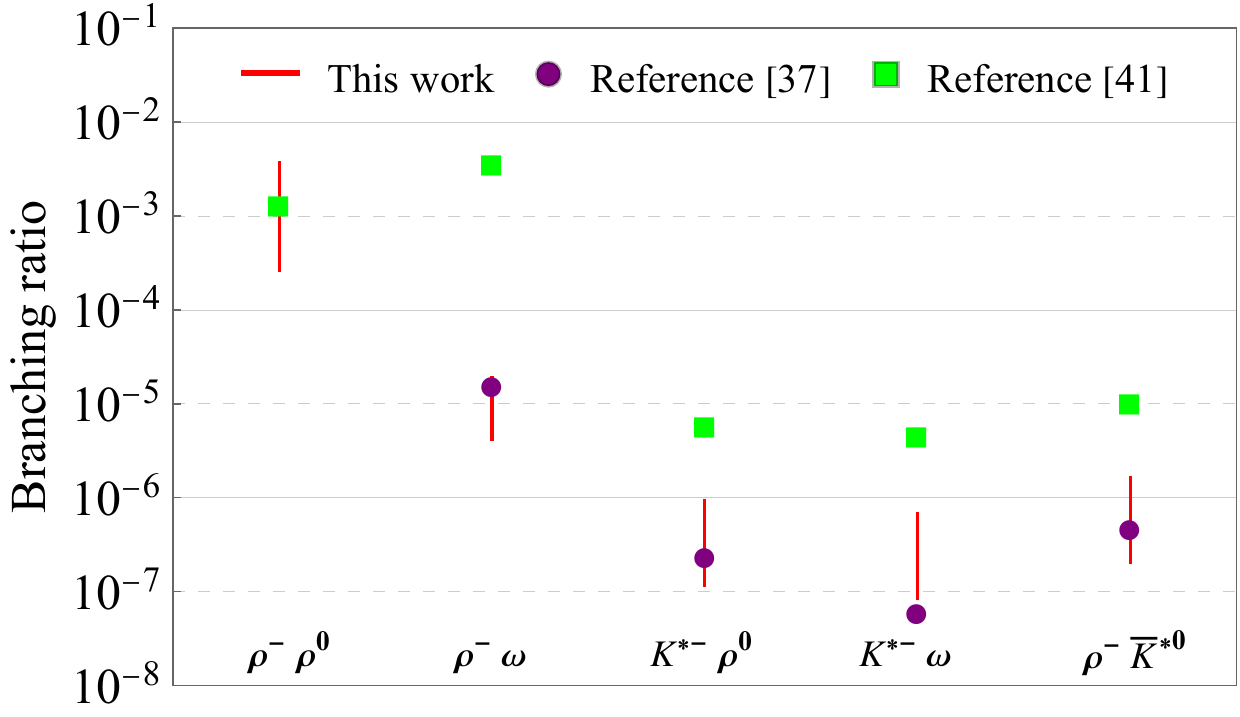}
  \end{minipage}
  \\
  \begin{minipage}{\linewidth}
    \centering
    \includegraphics[width=\linewidth]{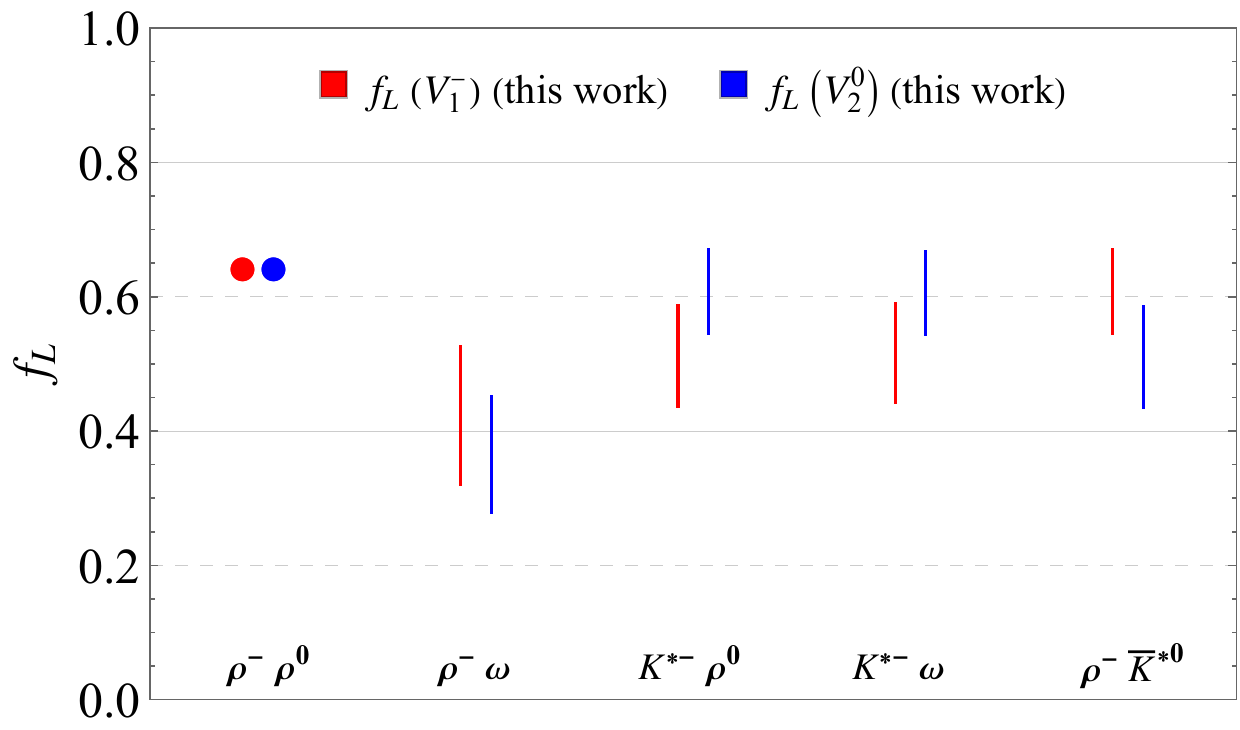}
  \end{minipage}
\caption{(Top) Branching fractions and $f_L$ for $ \tau^-\to V_1^-V_2^0\nu_\tau$ decays obtained in this work. For comparison, predictions of the branching ratios in the effective chiral model \cite{Li:1996ks} and the angular momentum algebra method \cite{Dai:2018thd} are also exhibited, where the one for $\tau^-\to \rho^-\rho^0\nu_\tau$ is not given in the former reference. (Bottom) Longitudinal polarization fractions in this work. For both top and bottom figures, red or blue lines represent regions that are obtained by varying the parameters, where the details are given in the texts. For $f_L$ in $\tau^-\to \rho^-\rho^0\nu_\tau$ decays, the varied ranges are rather tiny, of $\mathcal{O}(10^{-3})$, and the values averaged over the regions are listed for both $f_{L}(\rho^-)$ and $f_{L}(\rho^0)$}
\label{Fig:3}
\end{figure}
\noindent
\bea
T^{(A)}_{\rm GHLS}\propto (p_1-p_2)^1,
\quad
T^{(A)}_{\rm TAUOLA}\propto (p_1-p_2)^0.\label{Eq:comparison}
\eea 
Due to Eq.~(\ref{Eq:comparison}), the latter typically gives a larger contribution since the former is phase-space suppressed. 
\par
If we adopt $T^{(A)}_{\rm TAUOLA}$ with $c_0=3$ \cite{Kuhn:2006nw} in Eq.~(\ref{Eq:Tauoexp1}), this dominates the branching ratio over other contributions, and results in $\textrm{Br}[\tau^-\to \rho^-\omega\nu_\tau]=\mathcal{O}(1)\%$, an order of magnitude larger than the result \cite{Dai:2018thd} shown in Fig.~\ref{Fig:3}. In view of this aspect, we take GHLS as a primary case to evaluate the lowest-lying contribution in
Fig.~\ref{Fig:2}(b).
\par
Furthermore, for CS channels,  $K_{1A}$ in Fig.~2(b) represents the axial-vector meson that constitutes $J^{PC}=1^{++}$ nonet with $a_1(1260)$. Mixing between $K_{1A}$ and $K_{1B}$, where the latter is a component of the $1^{+-}$ nonet, occurs in the presence of flavor symmetry breaking \cite{Cheng:2011pb,Dai:2020vfc,Hayasaka:2021ecj}:
\bea
\begin{pmatrix}
\ket{K_1(1270)}\\
\ket{K_1(1400)}
\end{pmatrix}=
\begin{pmatrix}
\sin\theta_{K_1} & \cos\theta_{K_1}\\
\cos\theta_{K_1} &
-\sin\theta_{K_1}
\end{pmatrix}
\begin{pmatrix}
\ket{K_{1A}}\\
\ket{K_{1B}}
\end{pmatrix},
\eea
with favored values of $\theta_{K_1}$ given by $34^\circ$ and $57^\circ$ \cite{Cheng:2011pb}, both of which are considered in the later numerical results.
\par
The lowest-lying contributions in Fig.~\ref{Fig:2}(a), and (c), and also nonresonant one in (d), include parameters in HLS, $g, a$, and $c_3$. Relevant values of these  quantities in Ref.~\cite{Harada:2003jx} are $g=6.00$, $c_3=0.61$, $a=2.07$ (see also Refs.~\cite{Bramon:1994pq,Hashimoto:1996ny,Harada:2011xx} for discussion to determine the pamameters in HLS), used in the later analysis.
\par
For the excited-state (denoted here by $M^\prime$) contributions, couplings for $V_1V_2M^\prime$ and $WM^\prime$ are modeled by interactions that have Lorentz structure identical to the ones of (G)HLS in Eqs.~(\ref{Eq:DINT1})-(\ref{Eq:DINT4}) for the corresponding lowest-lying states. Regarding determinations of these couplings, the relevant procedure is given as follows: for $\rho\rho\rho(1700)$, widths of $\Gamma(\rho(1700)\to \rho\rho)$ in Refs.~\cite{He:2013ttg, Barnes:1996ff} are used to fix the range of $G_{\rho\rho\rho(1700)}$. With this fixed parameter, the $\rho\rho\rho(1450)$ coupling is determined within the range of $G_{\rho\rho\rho(1700)}\leq G_{\rho\rho\rho(1450)}\leq G_{\rho\rho\rho}|_{\rm HLS}$. The branching ratio, $f_L(\rho^-),$ and $f_L(\rho^0)$ in $\tau^-\to \rho^-\rho^0\nu_\tau$ decays are obtained by scanning over possible ranges of these parameters. As for the $WM^\prime$ couplings, the details are given around Eqs.~(\ref{Eq:GWrho})-(\ref{Eq:GWV3}), which are applied also for other $\tau^-\to V_1^-V_2^0\nu_\tau$ decays.
\par
In the case of $\tau^-\to \rho^-\omega\nu_\tau$, contributions of intermediate excited states of $a_1$ and $\pi$ are present. The couplings  of $\rho\omega a_1(1640)$ and $\rho\omega\pi(1800)$ are determined by $\Gamma(a_1(1640)\to \rho\omega)$ \cite{Chen:2015iqa} and $\Gamma(\pi(1800)\to \rho\omega)$ \cite{Barnes:1996ff}, respectively, in a way similar to the previous procedure for $\rho\rho\rho(1700)$, and a range for the $\rho\omega\pi(1300)$ coupling is obtained as $G_{\rho\omega\pi(1800)}\leq G_{\rho\omega\pi(1300)}\leq G_{\rho\omega \pi}|_{\rm HLS}$. There is an additional undetermined parameter for the lowest-lying $\rho\omega a_1(1260)$ coupling in GHLS, especially $\tilde{c}_8$ defined in Eq.~(\ref{Eq:c8tildeeq}). This coefficient can be fixed by charmed meson decay in the large-$N_c$ limit via $\textrm{Br}[D_s^+\to \rho^+\omega]=0.99\times 10^{-2}$ \cite{BESIII:2025owp}: Including $a_1(1260), a_1(1640), \pi$, and the nonresonant contribution, $D_s^+\to \rho^+\omega$ is analyzed, and $\tilde{c}_8$ is determined within twofold ambiguity, under the fixed value of $G_{\rho\omega a_1(1640)}$. The procedure for evaluating $\textrm{Br}[D_s^+\to \rho^+\omega]$ is summarized in Ref.~\cite{SupplementalMaterial} while input decay constant and Wilson coefficients for the $D_s^+$ decay are extracted from Refs.~\cite{FlavourLatticeAveragingGroupFLAG:2024oxs,Buchalla:1995vs}. The obtained possible twofold values or ranges of couplings are scanned to obtain ranges for the branching ratio, $f_L(\rho^-)$, and $f_L(\omega)$, in $\tau^-\to \rho^-\omega\nu_\tau$ decays. 
\par
For the case of $\tau^-\to K^{*-} \rho^0\nu_\tau$, $\tau^-\to K^{*-} \omega\nu_\tau$ and $\tau^-\to \rho^- \bar{K}^{*0}\nu_\tau$, the similar procedure is implemented: a range for the coupling $G_{K^*(1680)}$ related to $G_{K^*\rho K^*(1680)}$, $G_{K^*\omega K^*(1680)}$, and $G_{\rho K^* K^*(1680)}$ via Eq.~(\ref{Eq:GV1V2coef}) is determined from $\Gamma(K^*(1680)\to K^*\rho)$ and $\Gamma(K^*(1680)\to K^*\omega)$ \cite{Pang:2017dlw}. Subsequently, the coupling corresponding to $K^*(1410)$ is obtained within the range of $G_{K^*(1680)}\leq G_{K^*(1410)}\leq \left.G_{K^*(892)}\right|_{\rm HLS}$. Combined with the range of the excited pseudoscalar coupling, $0\leq G_{K(1460)}\leq \left.G_{K}\right|_{\rm HLS}$, the observables are scanned over those parameter regions, with twofold variation of $\theta_{K_1}=34^\circ$ and $57^\circ$. For the mentioned resonant couplings, input widths and some of determined parameter ranges are summarized in Appendix E.
\par
In Fig.~\ref{Fig:3}, numerical results for the branching ratios and $f_L$ are exhibited, where the former quantities are compared with those from the previous works \cite{Li:1996ks, Dai:2018thd}. For the latter reference, the displayed branching ratios are extracted from Tab.~VI \cite{Dai:2018thd}, corresponding to the dominant $s$-wave without convolution, which is to be compared with other results. As seen from the figure, for $\tau^-\to\rho^-\rho^0\nu_\tau$, the obtained range covers the result of the angular momentum algebra model \cite{Dai:2018thd}. Regarding the channels other than $\tau^-\to\rho^-\rho^0\nu_\tau$, the size of the branching fractions in this work is comparable with the ones in the effective chiral model \cite{Li:1996ks}, while being smaller than those in Ref. \cite{Dai:2018thd}. 
\par
\begin{figure}[t]
\centering
\includegraphics[width=1.0\linewidth]{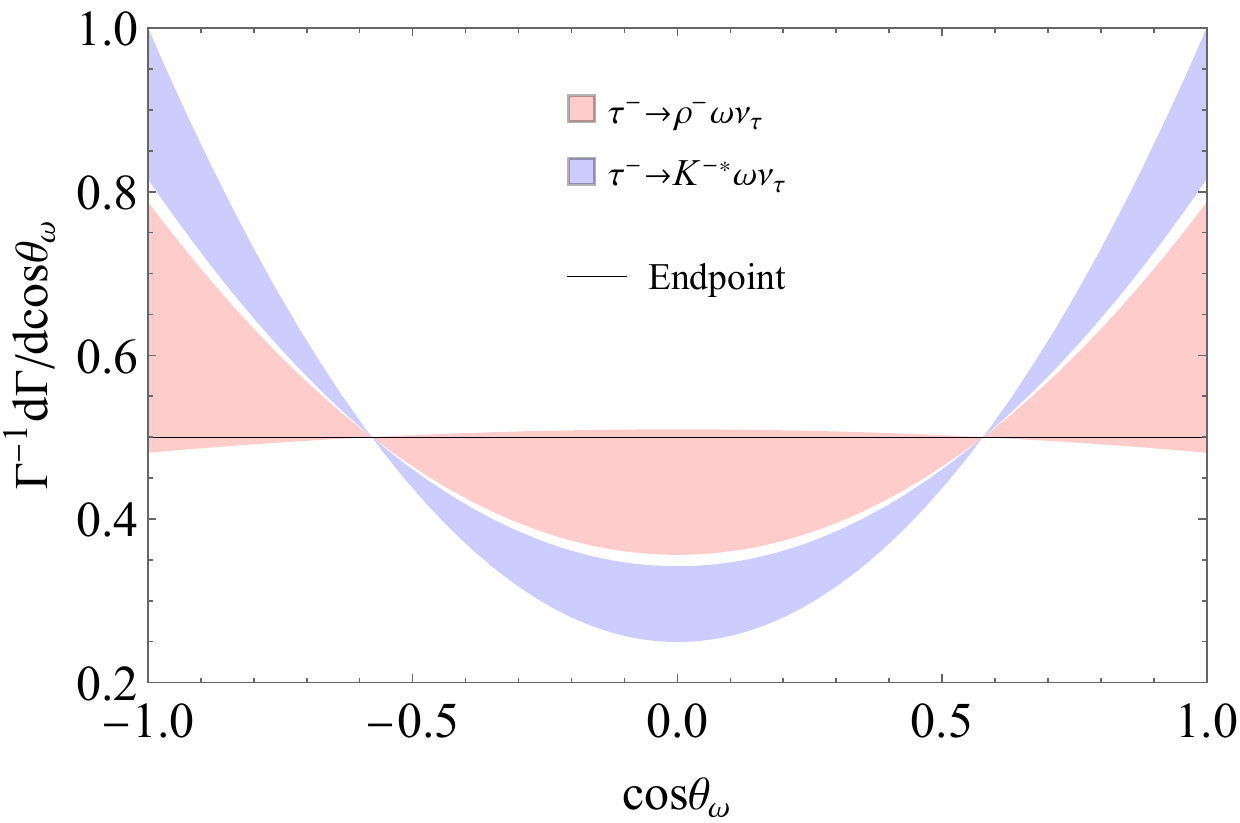}
\vspace{-8mm}
\caption{Angular distributions in $\tau^-\to V^-\omega\nu_\tau~(V=\rho, K^*)$ decays for selected cases. 
The bands represent the regions of $f_L$ in the bottom Fig.~\ref{Fig:3}. The black horizontal line corresponds to the prediction from the endpoint limit.}
\label{Fig:5}
\end{figure}

As for the longitudinal polarization fractions, the decay modes including Fig.~\ref{Fig:2}(a) give $0.44 \leq f_L \leq 0.67$, which tend to be larger than $0.28 \leq f_L \leq 0.53$ for $\tau^-\to \rho^-\omega\nu_\tau$ with no intermediate vector-meson contributions. In Fig.~\ref{Fig:5}, angular distributions are presented for some selected cases. Those results show that the distribution in $\tau^-\to \rho^-\omega\nu_\tau$ is different from others, due to the absence of Fig.~\ref{Fig:2}(a), compared with $\tau^-\to K^{-*}\omega\nu_\tau$ as an example.
\par
The longitudinal polarization fractions can be also analyzed by the TAUOLA-based interaction in Eqs.~(\ref{Eq:Tauoexp1}) and (\ref{Eq:Tauoexp2}) instead of (G)HLS. For this case, the dominant diagram in Fig.~\ref{Fig:2}(b) from $a_1(1260)$ leads to $f_L^{(\rho^-)}\simeq f_L^{(\omega)}= 0.34$ in $\tau^-\to \rho^-\omega\nu_\tau$ decays. However, these are not regarded as our primary results due to the large $\textrm{Br}(\tau^-\to \rho^-\omega\nu_\tau)$ and the absence of interaction structure in Eq.~(\ref{Eq:Tauoexp2}) for the case of GHLS, as mentioned before.
\par
The numerical results presented above are based on the case where width distributions of vector mesons in the final state are not considered (see Ref.~\cite{Dai:2018thd} for an analysis of branching ratios with nonvanishing widths). Another issue is that in channels such as $\tau^-\to \rho^- \pi^+\pi^-\nu_\tau$ decays, $\rho-\omega$ interference occurs, as is well known for $e^+e^-\to \pi^+\pi^-$, resulting in a more involved line shape of $\pi^+\pi^-$ invariant mass. Those aspects are beyond the scope of the present work.
\\
\\
\textit{Endpoint relations.} For $\tau^- \to V_1^- V_2^0 \nu_\tau$ decays, the smallest value of kinematically allowed invariant mass squared is given by $Q^2_{\rm min}=(m_{V_1}+m_{V_2})^2$. For this case, the produced vector mesons are not boosted in the rest frame of the hadronic system, {\it i.e.,} $|\bm{p}|=0$. At the endpoint, kinematical relations that follow from Lorentz symmetry were discussed in Ref.~\cite{Hiller:2013cza}. One of the consequences is that, taking $B\to K^*\ell^+\ell^-$ decay as an example, $f_L$ for a vector meson at the endpoint is fixed to a constant ($q=p_{\ell^+}+p_{\ell^-}$) \cite{Hiller:2013cza},
\bea
\left.f_L(q^2_{\rm max})\right|_{B\to K^*\ell^+\ell^-}=1/3.\label{Eq:endpBFCNC}
\eea
\vspace{-4mm}
\par
Discussions of the kinematical endpoints were also given for $D\to V_1V_2$ decays \cite{Hiller:2013cza}, in which case phase space region is limited due to $m_D\gtrapprox m_{V_1}+m_{V_2}$. Since this configuration is approximately close to the endpoint, $f_L\approx 1/3$ is expected in a way similar to Eq.~(\ref{Eq:endpBFCNC}), if the approximation of $m_D\simeq m_{V_1}+m_{V_2}$ works properly. However, $f_L(D^0\to \rho^0\rho^0)=0.71\pm 0.04\pm 0.02$ in FOCUS \cite{FOCUS:2007ern}, $f_L(D^0\to \omega\phi)<0.24$ at the 95$\%$ confidence level in BESIII \cite{BESIII:2021raf}, and $f_L(D^0\to K^{*+}K^{*-})=0.468\pm 0.046\pm 0.011$ in recent BESIII \cite{BESIII:2026dwz} are significantly deviated from the endpoint value, while $f_L(D^0\to K^{*-}\rho^+)=0.475\pm 0.271$ in MARK-III \cite{MARK-III:1991fvi} (see Ref.~\cite{Cheng:2024hdo}) is consistent within large uncertainty. Those results  imply that the endpoint approximation is inadequate for nonleptonic charm decays.
\par
In order to quantify how a kinematical configuration is close to an endpoint, we introduce \cite{Hiller:2013cza},
\bea
u_\tau(Q^2)=\frac{(m_{V_1}+m_{V_2})^2}{Q^2},\quad
u_P=\frac{(m_{V_1}+m_{V_2})^2}{m_P^2},\quad
\label{Eq:udefinitions}
\eea
associated with $\tau^-\to V_1^-V_2^0\nu_\tau$ and $P^-\to V_1^-V_2^0$ for $P=D, B$, respectively. Its domain is $0\leq u_f\leq 1~(f=\tau, D, B)$, with $u_f=1$ corresponding to an endpoint where vector mesons are not boosted in the rest frame of the hadronic system. For the three-body $\tau$ decays, $u_\tau(Q_{\rm min}^2)=1$ by definition while $u_\tau(Q_{\rm max}^2)$ with $Q_{\rm max}^2=m_\tau^2$ represents the maximal deviation from unity in kinematically allowed ranges. The values of those parameters are listed in Tab.~\ref{Tab:1}. For each of the considered channels, $u_B\ll u_D<u_\tau(Q_{\rm max}^2)$ is realized, as it should be. In what follows, $\tau$ and $D$ decays are discussed, which lie closer to the endpoints than $B\to V_1V_2$ decays.
\par
For $\tau^-\to V_1^-V_2^0\nu_\tau$ decays,
endpoint relations analogous to Eq.~(\ref{Eq:endpBFCNC}) are also satisfied {\it exactly} at $Q^2=Q^2_{\rm min}$, which can be confirmed straightforwardly: If $Q^2$ is set to $Q^2_{\rm min}$ in the entire phase space region of the integrand in Eqs.~(\ref{Eq:Keqationsec}) and (\ref{Eq:expresfuncI}), $f_L$ in Eq.~(\ref{Eq:FLdefrelations}) leads to
\bea
\left.f_L^{(V^-_1)}\right|_{Q^2\to Q^2_{\rm min}}
=\left.f_L^{(V_2^0)}\right|_{Q^2\to Q^2_{\rm min}}
=1/3.\label{Eq:appro}
\eea
We stress that although Eq.~(\ref{Eq:appro}) is consistent with Ref.~\cite{Hiller:2013cza}, practically nonnegligible corrections to $f_L=1/3$ are possible. Since $f_L$ is defined as an integral over $Q^2_{\rm min}\leq Q^2\leq Q^2_{\rm max}$, deviation from the endpoint limit occurs due to the region aside from $Q^2=Q^2_{\rm min}$. As shown in Fig.~\ref{Fig:3}, $f_L\neq 1/3$ can be realized for all five channels considered.
\par
From kinematical consideration, there is no wonder that $f_L(D\to V_1V_2)$ is deviated from $1/3$. This is because $D\to V_1V_2$ kinematics sits slightly away from the endpoint, as compared to $\tau\to V_1V_2\nu_\tau$ decays, where the latter is {\it already} deviated from $f_L=1/3$ as was obtained in this work. In this context, experimental verifications of $f_L$ for $\tau$ decays in Fig.~\ref{Fig:3} are of certain relevance in the future study.
\begin{table}[t]
\caption{Values of endpoint variables defined in Eq.~(\ref{Eq:udefinitions}) for $\tau^-\to V^-_1 V_2^0\nu_\tau$, $D^-\to V^-_1 V_2^0$, and $B^-\to V^-_1 V_2^0$ decays.}
\label{Tab:1}
\vspace{-3mm}
\begin{center}
\begin{tabular}{cccccc}\hline\hline
$V_1^-V_2^0$
& $\rho^-\rho^0$
& $\rho^-\omega$
& $K^{*-}\rho^0$
& $K^{*-}\omega$
& $\rho^{-}\bar{K}^{*0}$
\\\hline 
$u_{\tau}(Q_{\rm max}^2)$
&$0.76$
&$0.77$
&$0.88$
&$0.89$
&$0.88$
\\\hline 
$u_{D}$
&$0.69$
&$0.69$
&$0.80$
&$0.80$
&$0.80$
\\\hline 
$u_{B}$
&$0.09$
&$0.09$
&$0.10$
&$0.10$
&$0.10$
\\\hline \hline  
\end{tabular}
\end{center}
\vspace{-8mm}
\end{table}
\\
\\
\textit{Summary and outlook.} We have analyzed $\tau^-\to V_1^-V_2^0\nu_\tau$ decays to extract the polarized observables for vector mesons. The longitudinal polarization fractions denoted as $f_L$ are evaluated by GHLS Lagrangian for the lowest-lying particles, and the contributions of excited intermediate mesons are additionally included. The numerical results in the present work have shown that $f_L\neq 1/3$ are generically possible. This indicates that corrections from $|\bm{p}|\neq 0$ ($\bm{p}$ is a momentum of either vector meson in the hadronic rest frame) are sizable, since $f_L= 1/3$ is realized at the kinematical endpoint ($|\bm{p}|=0$). Meanwhile, for $D\to V_1V_2$ decays, several experimental data \cite{FOCUS:2007ern,BESIII:2021raf,BESIII:2026dwz} are significantly deviated from $f_L= 1/3$. These results indicate that use of the endpoint approximation is invalid at the scale of $m_\tau \sim m_D$. In light of this status, future measurement at Belle II constitutes a key step toward unraveling the patterns of the observables. Prospected in this way, $\tau^-\to V_1^-V_2^0\nu_\tau$ would offer a novel testing ground for polarizations, in attempt to obtain unified understanding of strong interaction at scales of $\tau$-charm physics, at the same time providing an insight on nonleptonic charm decays. Another aspect to be revealed is the size of the branching ratios: Taking $\tau^-\to\rho^-\omega\nu_\tau$ as an example, the angular momentum algebra model \cite{Dai:2018thd} gives $\textrm{Br}[\tau^-\to\rho^-\omega\nu_\tau]=\mathcal{O}(10^{-3})$, much larger than $\mathcal{O}(10^{-5})$ in the effective chiral model \cite{Li:1996ks}, where our results are comparable with the latter. Furthermore, the contribution of the axial-vector meson is sensitive to the structure of $a_1\rho\omega$ coupling: The one adopted in TAUOLA \cite{Kuhn:2006nw} discussed for $\tau\to 5\pi\nu_\tau$ decays gives even larger branching fraction, $\textrm{Br}[\tau^-\to \rho^-\omega\nu_\tau]=\mathcal{O}(10^{-2})$. However, this type of interaction was not obtained in the GHLS approach \cite{Kaiser:1990yf}. Future measurements in flavor factories are required for clarifying this issue.
\\
\\
\noindent
\textit{Acknowledgments.} The authors would like to thank Shinya Matsuzaki for answering a question on anomalous couplings in HLS. This work is supported by the National Science Foundation of China under Grant No.~12405111 for H.U. and the Seeds Funding of Jilin University.
\\
\appendix
\setcounter{equation}{0}
\renewcommand{\theequation}{A\arabic{equation}}
\noindent
\textit{Appendix A: Process dependence of polarized configurations.} For channels including vector meson(s) in the final state, possible configurations of polarizations are different, depending on a specific decay mode. Here we consider HAs based on the $V-A$ current intended for the SM, taking $\tau$ and $D$ decays as examples. In $B$-meson decays, polarized configurations are similar to the ones for $D$ mesons.
\par
For each channel, HAs are listed as,
\begin{align}
&\tau\to V\nu_\tau:&
H_{\lambda}^{(m)V}=&\epsilon_\mu(m)\braket{V(\lambda)|
J^\mu_q|0},\label{Eq:HVfirst}\\
&\tau\to VP\nu_\tau:&
H_{\lambda}^{(m)VP}=&\epsilon_\mu(m)\braket{V(\lambda)P|J^\mu_q|0},\\
&\tau\to VV\nu_\tau:&H_{\lambda_1\lambda_2}^{(m)VV}=&\epsilon_\mu(m)\braket{V_1(\lambda_1)V_2(\lambda_2)|J^\mu_q|0},\label{Eq:thirdEQVV}\\
&D\to VP:&
H_{\lambda}^{VP}=&\braket{V(\lambda)P|\mathcal{H}_W|D},\\
&D\to VV:&
H_{\lambda_1\lambda_2}^{VV}
=&\braket{V_1(\lambda_1)V_2(\lambda_2)|\mathcal{H}_W|D},\label{Eq:HVlast}
\end{align}
where $\epsilon_\mu(m)$ and $J^\mu_q$ are introduced in Eq.~(\ref{Eq:heldefeq}), and $\mathcal{H}_W$ represents $\Delta C=1$ effective Hamiltonian. In the above relations,
$H$'s for $\tau$ decays are associated with hadronic HAs while the ones for $D$-meson decays correspond to the entire amplitude. Although $H_{\lambda_1\lambda_2}^{(m)VV}$ in Eq.~(\ref{Eq:thirdEQVV}) is just identical to the one in Eq.~(\ref{Eq:heldefeq}), it is introduced for the sake of definitiveness of the notations.
\par
In general, polarizations without any restrictions are
\bea
&\lambda=\pm, 0,&\label{Eq:Helconf1}\\
&(\lambda_1, \lambda_2)=(\pm, \pm), (\pm, \mp), (0, \pm), (\pm, 0), (0, 0).&
\label{Eq:Helconf2}
\eea
In $\tau\to V\nu_\tau$ (for unpolarized $\tau$) and $\tau\to VP\nu_\tau$ decays, the number of helicity configurations of the vector meson is three for both channels, in accordance with Eq.~(\ref{Eq:Helconf1}). However, for other cases, parts of helicity configurations vanish due to kinematical restrictions. This can be understood by expanding Eqs.~(\ref{Eq:HVfirst})-(\ref{Eq:HVlast}) in terms of all possible Lorentz structures, and substituting the relevant polarizations in Eqs.~(\ref{Eq:A1})-(\ref{Eq:A3}) and (\ref{Eq:Vpol1})-(\ref{Eq:Vpol2}). In $\tau\to VV\nu_\tau$ decays, it is straightforward to show that two helicity configurations vanish,
\bea
H^{(m)VV}_{\pm\mp}=0.\label{Eq:VVvanish}
\eea
We note that the relation in Eq.~(\ref{Eq:VVvanish}) is realized unless there is an additional contribution of a tensor particle that replaces $W$ boson. Due to Eqs.~(\ref{Eq:Helconf2}) and (\ref{Eq:VVvanish}), there exist seven nonvanishing hadronic HAs in general for $\tau\to VV\nu_\tau$ decays: Two modes where both $V_1$ and $V_2$ are transverse ($H_{\pm\pm}^{(m)VV}$),
one mode where both $V_1$ and $V_2$ are longitudinal ($H_{00}^{(m)VV}$),
and four modes where one is transverse while the other is longitudinal ($H_{\pm 0}^{(m)VV}$ and $H_{0\pm}^{(m)VV}$).
\par
In the case of $D$-meson decays, vanishing entries read,
\bea
H^{VP}_{\pm}=0,\quad H^{VV}_{\pm\mp}=H^{VV}_{0\pm}=H^{VV}_{\pm0}=0.\label{Eq:Dvanish}
\eea
Hence, the numbers of nonzero HAs are one and three for $D\to VP$ and $D\to VV$, respectively. The resulting numbers of polarization configurations are summarized in Tab.~\ref{Tab:2}.
\\
\begin{table}[t]
\begin{center}
\caption{Numbers of nonvanishing polarized configurations for vector meson(s) in $\tau$ decays from the $V-A$ current and $D$-meson decays. Unpolarized $\tau$ is considered for $\tau\to V\nu_\tau$.}
{\scriptsize
\label{Tab:2}
\begin{tabular}{cccccc}\hline\hline
& $\tau\to V\nu_\tau$ 
& $\tau\to VP\nu_\tau$
& $\tau\to VV\nu_\tau$
& $D\to VP$
& $D\to VV$
\\\hline 
$\#$ of pols.
& 3
& 3
& 7
& 1
& 3
\\\hline\hline
\end{tabular}
}
\end{center}
\vspace{-5mm}
\end{table}
\par
\setcounter{equation}{0}
\renewcommand{\theequation}{B\arabic{equation}}
\noindent
\textit{Appendix B: Polarization vectors.} Here conventions of polarization vectors used in the present work are summarized. For the one in Eq.~(\ref{Eq:heldefeq}), the following basis is adopted,
\bea
\epsilon^\mu(\pm)&=&
\frac{1}{\sqrt{2}}(0, \pm 1, i, 0),\label{Eq:A1}\\
\epsilon^\mu(0)&=&(0, 0, 0, -1),\\
\epsilon^\mu(s)&=&(1, 0, 0, 0),\label{Eq:A3}
\eea
satisfying the normalization condition,
\bea
g_{\mu\nu}\epsilon^\mu(m)
\epsilon^{*\nu}(m^\prime)
&=&g_{mm^\prime},
\eea
for $m, m^\prime=s, \pm, 0$. As for the two vector mesons in the final state of $\tau^-\to V_1^-V_2^0\nu_\tau$, we introduce the set of quantities,
\bea
\epsilon_{1}^{\mu}(\pm)&=&
\frac{1}{\sqrt{2}}(0, \pm 1, i, 0),\label{Eq:Vpol1}\\
\epsilon^{\mu}_1(0)&=&\frac{1}{m_{V_1}}(|\bm{p}|, 0, 0, E_{1}),\\
\epsilon^{\mu }_{2}(\pm)&=&
\frac{1}{\sqrt{2}}(0, \pm 1, -i, 0),\\
\epsilon^{\mu}_2(0)&=&\frac{1}{m_{V_2}}(|\bm{p}|, 0, 0, -E_2),\label{Eq:Vpol2}
\eea
for which the following relations hold
\bea
g_{\mu\nu}\epsilon^\mu_i(m)
\epsilon^{*\nu}_i(m^\prime)&=&-\delta_{mm^\prime},\label{Eq:pol22}\\
\delta_{mm^\prime}
\epsilon_{i}^\mu(m)
\epsilon_{i}^{\nu*}(m^\prime)&=&
-g^{\mu\nu}+\frac{p_{i}^\mu p_{i}^\nu}{m_{V_i}^2},\label{Eq:pol23}
\eea
with $m, m^\prime=\pm, 0$.
In Eqs~(\ref{Eq:pol22}), (\ref{Eq:pol23}), sum over $i=1, 2$ is not taken.
\par
\setcounter{equation}{0}
\renewcommand{\theequation}{C\arabic{equation}}
\noindent\textit{Appendix C: Angular observables.} Definitions for some objects introduced
in Eqs.~(\ref{Eq:ResAngfinalfin2}), (\ref{Eq:ResAngfinalfin3}), (\ref{Eq:FLdefrelations})
are given by \cite{SupplementalMaterial},
\bea
K&=&\displaystyle\sum_{\lambda_1,\:\lambda_2}K_{\lambda_1\lambda_2},
\label{Eq:Kequationfirst}
\\
K_{\lambda_1\lambda_2}&=&\int dQ^2
\frac{|\bm{p}|}{\sqrt{Q^2}}
\left(1-\frac{Q^2}{m_\tau^2}\right)^2I_{\lambda_1\lambda_2},\label{Eq:Keqationsec}\\
I_{\lambda_1\lambda_2}&=&
I_{\lambda_1\lambda_2}^{(+1/2)}
+\frac{m_\tau^2}{Q^2}
I_{\lambda_1\lambda_2}^{(-1/2)}.\label{Eq:expresfuncI}
\eea
In Eq.~(\ref{Eq:expresfuncI}), the indices of $\pm 1/2$ denote helicities of the initial $\tau$ lepton, and we introduced \cite{SupplementalMaterial}
\bea
I_{\lambda_1\lambda_2}
^{(+1/2)}
&=&2\left[S^{(++)}_{\lambda_1\lambda_2}+S^{(--)}_{\lambda_1\lambda_2}
+S^{(00)}_{\lambda_1\lambda_2}\right]
\nn\\
&+&2\textrm{Re}(S^{(+-)}_{\lambda_1\lambda_2})
-\frac{3\sqrt{2}\pi}{4}\textrm{Re}[S^{(+0)}_{\lambda_1\lambda_2}
+S^{(-0)}_{\lambda_1\lambda_2}],\nn\\
\label{Eq:Iexp1}
\\
I_{\lambda_1\lambda_2}
^{(-1/2)}
&=&S^{(++)}_{\lambda_1\lambda_2}+S^{(--)}_{\lambda_1\lambda_2}+S^{(00)}_{\lambda_1\lambda_2}+3S^{(ss)}_{\lambda_1\lambda_2}
\qquad\nn\\
&-&2\textrm{Re}(S^{(+-)}_{\lambda_1\lambda_2})
-\frac{3\sqrt{2}\pi}{4}
\textrm{Re}[S^{(+s)}_{\lambda_1\lambda_2}-S^{(-s)}_{\lambda_1\lambda_2}],\nn\\
\label{Eq:Iexp2}
\eea
which are based on the adopted convention in Eqs.~(\ref{Eq:A1})-(\ref{Eq:A3}),
with $S$ defined as an object that depends quadratically on the hadronic HA,
\bea
S^{(mm^\prime)}_{
\lambda_1\lambda_2}
=H_{\lambda_1\lambda_2}^{(m)}
H_{\lambda_1\lambda_2}^{(m^\prime)*}.
\eea
\par
\setcounter{equation}{0}
\renewcommand{\theequation}{D\arabic{equation}}
\noindent
\textit{Appendix D: Hadronic helicity amplitudes.} 
For the case where the intermediate meson ($M$) is a mass eigenstate, the coefficients of the HA defined in Eq.~(\ref{Eq:HVexpre1}) for the intermediate $V, A$, and $P$ contributions are decomposed by
\bea
C^{(i)}_{V_1V_2M}&=&G_{WM} G_{V_1V_2M} D_M,\label{Eq:vertCi}\\
D_M&=&\frac{1}{m^2_M-Q^2-i m_M\Gamma_M},
\eea
where in Eq.~(\ref{Eq:vertCi}), the $G_{WM}$ represents the coupling between $W$ boson and the intermediate resonance, while in Eqs.~(\ref{Eq:NRcoeff}) and (\ref{Eq:vertCi}), $G_{V_1V_2 M}$ denotes the $V_1V_2 M$ coupling. In (G)HLS, the expressions for the $WM$ couplings
at the lowest order are,
\bea
G_{W\rho}=G_{WK^*}&=&2\sqrt{2}gf_\pi^2,\label{Eq:GWrho}\\
G_{Wa_1(1260)}=G_{WK_{1A}}
&=& \sqrt{2} g  f_\pi^2,\\
G_{W\pi}=G_{WK}&=&f_\pi.\label{Eq:GWr}
\eea
Using the relations in HLS \cite{Harada:2003jx} and GHLS \cite{Kaiser:1990yf} such as
$m_\rho=\sqrt{2a}f_\pi g$,
$f_\rho=2f_\pi/\sqrt{a}$,
$m_{a_1}=\sqrt{2}g f_\pi$,
and $f_{a_1}=f_\pi$,
Eqs.~(\ref{Eq:GWrho})-(\ref{Eq:GWr}) are given by the forms
\bea
G_{WV}&=&f_Vm_V,\label{Eq:GWV1}\\
G_{WA}&=&f_Am_A,\\
G_{WP}&=&f_P.\label{Eq:GWV3}
\eea
The relations in Eqs.~(\ref{Eq:GWV1})-(\ref{Eq:GWV3}), corresponding to $WM$ coupling written as the onshell decay constants at the leading order, are used for excited-meson couplings to $W$ bosons. Input decay constants in the analysis are summarized in Tab.~\ref{Tab:dec}.
For the $K^*(1680)$ decay constant, the PDG assignment of its quantum number is $1^3D_1$, where the corresponding light unflavored vector meson is $\rho(1700)$, and SU(3) limit relation of $f_{K^*(1680)}\simeq f_{\rho(1700)}$ is used.
\begin{table}
\begin{center}
\caption{Decay constants in MeV units and references. For $K^*(1410)$, the branching ratio of $\tau$ decays \cite{ParticleDataGroup:2024cfk} is converted to the decay constant.
For $K(1460)$, difference between normalization of decay constants in this work and one in Ref.~\cite{Volkov:2019yhy} is considered properly.}
\label{Tab:dec}
\begin{tabular}{ccccccc}\hline\hline
$f_{\rho(1450)}$&
$f_{\rho(1700)}$&
$f_{K^*(1410)}$&
$f_{K^*(1680)}$&
$f_{a_1(1640)}$
\\\hline
$0.151$&
$0.14$&
$0.121$&
$0.14$&
$0.2249$
\\\hline
\cite{Jiang:2015paa}&
\cite{Li:2020ejs}&
\cite{ParticleDataGroup:2024cfk}&
\cite{Li:2020ejs}&
\cite{talk}
\\\hline\hline
\vspace{-2mm}
\\\hline\hline
$f_{\pi}$&
$f_{\pi(1300)}$&
$f_{\pi(1800)}$&
$f_{K(1460)}$&
\\\hline
$0.1302$&
$0.012$&
$0.0085$&
$0.042$&
\\\hline
\cite{FlavourLatticeAveragingGroupFLAG:2024oxs}&
\cite{Li:2021jqb}&
\cite{Li:2021jqb}&
\cite{Volkov:2019yhy}&
\\\hline\hline
\end{tabular}
\end{center}
\end{table}
\par
For the case where $M$ is a ground-state meson or $W$ boson, $G_{V_1V_2 M}$ in the (G)HLS is defined as vertices below,
\bea
\mathcal{L}_{V_1V_2V}&=&
iG_{V_1V_2V}
V^{-\mu}[
2V^{+\nu}_1(\partial_\nu V_{2\mu}^0)
-2(\partial_\nu V^{+}_{1\mu})V_2^{0\nu} \nn\\
&+&
(\partial_\mu V^{+}_{1\nu})V_2^{0\nu}
-V^{+\nu}_1(\partial_\mu V_{2\nu}^0)
],\label{Eq:DINT1}\\
\mathcal{L}_{V_1V_2A}&=&G_{V_1V_2 A}\epsilon^{\mu\nu\alpha\beta}
A_{\mu}^-\nn\\
&\times&[(\partial_\nu V_{1\alpha}^+)V_{2\beta}^0 -
V_{1\alpha}^+(\partial_\nu V_{2\beta}^0)],\label{Eq:DINT2}\\
\mathcal{L}_{V_1V_2P}&=&-G_{V_1V_2 P}P^-\epsilon^{\mu\nu\alpha\beta}\partial_\mu V_{1\nu}^{+} \partial_\alpha V_{2\beta}^{0},\label{Eq:DINT3}\\
\mathcal{L}_{V_1V_2W}&=&-\frac{g_2V_{uq}^*}{2\sqrt{2}}
G_{V_1V_2 W}\epsilon^{\mu\nu\alpha\beta}
W_\mu^-\nn\\
&\times&[(\partial_\nu V_{1\alpha}^{+})V_{2\beta}^0
+V_{1\nu}^{+}(\partial_\alpha  V_{2\beta}^0)].\label{Eq:DINT4}
\eea
The coupling is further decomposed by 
\bea
G_{V_1V_2 M}&=&N_{V_1V_2}^{(i)}G_{M}
~~(i=V,  A,  P,  \textrm{NR}).\label{Eq:GV1V2coef}
\eea
In the above relation, $N^{(i)}_{V_1V_2}$ is the normalization factor determined by SU(3) relations, summarized in Tab.~\ref{Tab:3}. The coupling coefficients in Eq.~(\ref{Eq:GV1V2coef}) in (G)HLS read ($N_c=3$),
\bea
\left. G_{\rho^-(770)}\right|_{\rm HLS}&=&\left. G_{K^{*-}(892)}\right|_{\rm HLS}=g,\label{Eq:COUPLING1}\\
\left. G_{a^-_1(1260)}\right|_{\rm GHLS}&=&\left. G_{K^-_A}\right|_{\rm GHLS}=-2g^2 \tilde{c}_8,\label{Eq:c8tildeeq}\\
G_{\pi^-}|_{\rm HLS}&=&G_{K^-}|_{\rm HLS}=\frac{N_cg^2c_3}{4\sqrt{2}\pi^2 f_\pi},\\
\left. G_{\rm NR}\right|_{\rm HLS}&=&\frac{N_cg^2 c_3}{8\sqrt{2}\pi^2}.\label{Eq:COUPLING4}
\eea
In Eqs.~(\ref{Eq:COUPLING1})-(\ref{Eq:COUPLING4}), $g$ and $c_3$ ($\tilde{c}_8$) are parameters in HLS \cite{Harada:2003jx} (GHLS \cite{Kaiser:1990yf}) Lagrangian. As for the Lorentz-scalar objects in Eqs.~(\ref{Eq:HVexpre1})-(\ref{Eq:NRcoeff}), expressions are given by ($\epsilon^{\mu\nu\rho\sigma}$ represents totally antisymmetric tensor)
\bea
T^{(V)}_{\cal V}&=&
\epsilon_\mu\nn
\left[2(p_2\cdot \epsilon_{1}^*)\epsilon_{2}^{*\mu}
-2(p_{1}\cdot \epsilon_{2}^*)
\epsilon_{1}^{*\mu}\right.\nn\\
&+&(\epsilon_{1}^*\cdot \epsilon_2^{*})
(p_{1}-p_{2})^\mu\nn\\
&-&
\left.
\frac{m_{V_1}^2-m_{V_2}^2}{m_{\cal V}^2}
(\epsilon_{1}^*\cdot \epsilon_{2}^*)Q^\mu
\right],\label{Eq:aGHLS}\\
T^{(A)}_{\cal A}&=&-T^{(\textrm{NR})}+\frac{2}{m_{\cal A}^2}T^{(P)},
\label{Eq:AxialGHLS}\\
T^{(P)}&=&
\frac{i}{2}(\epsilon\cdot Q) 
\epsilon^{\mu\nu\alpha\beta}
Q_\mu (p_1-p_2)_\nu
\epsilon_{1 \alpha}^{*}
\epsilon_{2 \beta}^{*},\\
T^{(\textrm{NR})}&=&
i\epsilon^{\mu\nu\alpha\beta}
\epsilon_\mu (p_1-p_2)_\nu
\epsilon_{1 \alpha}^{*}
\epsilon_{2 \beta}^{*}.
\label{Eq:texpfinal}
\eea
\begin{table}[t]
\begin{center}
\caption{Normalization factors of HAs defined in Eq.~(\ref{Eq:GV1V2coef}). Vanishing entries follow from $G$-parity conservation. Although $\tau^-\to K^{*-}K^{*0}\nu_\tau$ decays are kinematically prohibited unless resonance tails of vector mesons are considered, the coefficients are listed here. The last row gives whether a channel is CKM favored or suppressed.}
\label{Tab:3}
\begin{tabular}{ccccccc}\hline\hline
$V_1^-V_2^0$
& $\rho^-\rho$ 
& $\rho^-\omega$
& $K^{*-}K^{*0}$
& $K^{*-}\rho$
& $K^{*-}\omega$
& $\rho^{-}\bar{K}^{0*}$
\\\hline 
$N^{(V)}_{V_1^-V^0_2}$ 
& $1$
& $0$
& $1/\sqrt{2}$  
& $1/2$
& $1/2$
& $1/\sqrt{2}$
\\\hline
$N^{(A)}_{V_1^-V^0_2}$ 
& $0$
& $1$
& $1/\sqrt{2}$
& $1/2$
& $1/2$
& $1/\sqrt{2}$
\\\hline
$N^{(P)}_{V_1^-V^0_2}$ 
& $0$
& $1$
& $1/\sqrt{2}$
& $1/2$
& $1/2$
& $1/\sqrt{2}$
\\\hline
$N^{(\textrm{NR})}_{V_1^-V^0_2}$ 
& $0$
& $1$
& $1/\sqrt{2}$ 
& $1/2$
& $1/2$
& $1/\sqrt{2}$ 
\\\hline
CKM
& CF
& CF
& CF
& CS
& CS
& CS 
\\\hline
\hline
\end{tabular}
\end{center}
\end{table}
\par
Meanwhile, as discussed in the main texts, the contribution of an axial-vector meson for $\tau^-\to \rho^-\omega\nu_\tau$ can be extracted from the analysis of $\tau\to 5\pi\nu_\tau$ in Ref.~\cite{Kuhn:2006nw},
\bea
\left.C^{(A)}_{\rho^-\omega a_1(1260)}\right|_{\textrm{TAUOLA}}
&=&c_0 m_{a_1}^2
D_{a_1},\label{Eq:Tauoexp1}\\
\left.T^{(A)}\right|_{\textrm{TAUOLA}}
&=&-i\epsilon^{\mu\nu\alpha\beta}
\epsilon_\mu
\epsilon_{1 \nu}
\epsilon_{2 \alpha}
Q_\beta,\qquad\label{Eq:Tauoexp2}
\eea
to be compared with the GHLS results: Although the coefficient in Eq.~(\ref{Eq:Tauoexp1}) is just related to the one in Eq.~(\ref{Eq:vertCi}) for $(V_1, V_2, i, M)=(\rho^-, \omega, A, a_1(1260))$ by redefining the overall parameter, momentum dependence of Eq.~(\ref{Eq:Tauoexp2}) is different from the GHLS one in Eq.~(\ref{Eq:AxialGHLS}).
\par
For the $K_{1A}$ contribution, the relevant expression of Eq.~(\ref{Eq:HVexpre2}) is written as, 
\bea
\left.\mathcal{H}^{(A)}_{V_1V_2}\right|_{K_{1A}}&=&
\sin^2\theta_{K_1}
G_{WK_{1A}} 
G_{V_1V_2K_{1A}}
\nn\\
&\times&
D_{K_{1}(1270)}
T^{(A)}_{K_{1}(1270)}
\nn\\
&+&
\cos^2\theta_{K_1}
G_{WK_{1A}} 
G_{V_1V_2K_{1A}}\nn\\
&\times&
D_{K_1(1400)}
T^{(A)}_{K_1(1400)}.
\eea
\setcounter{equation}{0}
\renewcommand{\theequation}{E\arabic{equation}}
\textit{Appendix E: Couplings between an excited meson and two lowest-lying vector mesons.} The three-point-vertex-coefficients can be extracted by width of the excited meson decays,
\bea
\Gamma[\rho(1700)\to \rho\rho]&=&14~\textrm{MeV}~\textrm{\cite{Barnes:1996ff}},
\label{Eq:AppE2}\\
\Gamma[\rho(1700)\to \rho\rho]&=&  105~\textrm{MeV}~\textrm{\cite{He:2013ttg}},
\label{Eq:AppE3}\\
\Gamma[a_1(1640)\to \rho\omega]&=&[4,  18]~\textrm{MeV}~\textrm{\cite{Chen:2015iqa}},
\label{Eq:AppE1}\\
\Gamma[K^*(1680) \to K^*\rho]&=&6.33~\textrm{MeV~\cite{Pang:2017dlw}},
\label{Eq:AppE4}\\
\Gamma[K^*(1680) \to K^*\omega]&=&1.69~\textrm{MeV~\cite{Pang:2017dlw}},\label{Eq:AppE5}\\
\Gamma[\pi(1800)\to \rho\omega]&=&74~\textrm{MeV}~\textrm{\cite{Barnes:1996ff}},\label{Eq:AppE6}
\eea
which are converted into,
\bea
G_{\rho\rho\rho(1700)}&=&[0.67,  1.8],\\
G_{\rho\omega a_1(1640)}&=&[0.56,  1.2],\\
G_{K^*(1680)}&=&[0.49,  0.91],\\
G_{\rho\omega\pi(1800)}&=&3.1 ~\textrm{GeV}^{-1}.
\eea
In the those relations, the ranges of $G_{\rho\rho \rho(1700)}$, $G_{\rho\omega a_1(1640)}$, and $G_{K^*(1680)}$ are from Eqs.~(\ref{Eq:AppE2}) and (\ref{Eq:AppE3}), Eq.~(\ref{Eq:AppE1}), and Eqs.~(\ref{Eq:AppE4}) and (\ref{Eq:AppE5}), respectively, while $G_{\rho\omega\pi(1800)}$ is extracted from Eq.~(\ref{Eq:AppE6}).

\end{document}